\documentclass[twocolumn,prl,showpacs]{revtex4}

\usepackage{graphicx}
\usepackage{dcolumn}%
\usepackage{bm}%
\usepackage{amssymb}

\def\be{\begin{equation}}
\def\ee{\end{equation}}

\def\r{{\bm{r}}}

\begin{document}

\title{Columnar and lamellar phases in attractive colloidal systems}
\author{A. de Candia$^{a,b,c}$, E. Del Gado$^{a,b}$, A. Fierro$^{a,d}$,
N. Sator$^{e}$, M. Tarzia$^{a,b}$, and A. Coniglio$^{a,d}$}
\affiliation{${}^a$ Dipartimento di Scienze Fisiche, Universit\`a di
Napoli ``Federico II'',\\ Complesso Universitario di Monte
Sant'Angelo, via Cintia 80126 Napoli, Italy} \affiliation{${}^b$
CNISM Universit\`a di Napoli ``Federico II''}
\affiliation{${}^c$ INFN Udr di Napoli}
\affiliation{${}^d$ INFM CNR Coherentia}
\affiliation{${}^e$ Laboratoire de Physique Th\'eorique de la
Mati\`ere Condens\'ee, Universit\'e Pierre et Marie Curie-Paris 6 UMR
(CNRS) 7600 Case 121, 4 Place Jussieu 75252 Paris Cedex 05, France}
\date{\today}
\begin{abstract}
In colloidal suspensions, at low volume fraction and temperature, dynamical arrest occurs
via the growth of elongated
structures, that aggregate to form a connected network at gelation. Here we show that,
in the region of parameter space where gelation occurs, the stable thermodynamical phase
is a crystalline columnar one. Near and above the gelation threshold, the disordered spanning
network slowly evolves and finally orders to form the crystalline structure.
At higher volume fractions the stable phase is a lamellar one, that seems to have a still longer
ordering time.
\end{abstract}


\maketitle

In colloidal suspensions solid (or liquid) mesoscopic particles are 
dispersed in another substance. These systems, like blood, proteins in water, 
milk, black ink or paints, are important in our everyday lives, in biology 
and industry \cite{russel,morrison}. 
It is crucial, for example, to control the process of aggregation in paint 
and paper industries \cite{lu}, or to favour the protein crystallization 
in the production of pharmaceuticals and photonic crystals 
\cite{muschol,mcpherson}.

A practical and exciting feature of colloidal suspensions is that
the interaction energy between particles can be well controlled
\cite{pusey,stradner,anderson}. In fact particles can be coated and stabilized
leading to a hard sphere behaviour, and an attractive depletion
interaction can be brought out by adding some non-adsorbing
polymers. The range and strength of the potential are controlled respectively 
by the size and concentration of the polymer \cite{anderson,edinburgh}.
Recent experimental works highlighted the presence of a net
charge on colloidal particles \cite{stradner,bartlett} giving rise to
a long range electrostatic repulsion in addition to the depletion
attraction.

The competition between attractive and repulsive interactions produces
a rich phenomenology and a complex behavior as far as structural
and dynamical properties are concerned. For particular choices
of the interaction parameters, the aggregation of particles is favoured but 
the liquid-gas phase transition can be avoided and the cluster size can 
be stabilized at an optimum value\cite{gron}. 
Experimentally, such a cluster phase made of small
equilibrium monodisperse clusters is observed using confocal microscopy at 
low volume fraction and low temperature (or high attraction strength) 
\cite{weitzclu,bartlett,stradner}. Increasing the volume fraction, 
the system is transformed from an ergodic cluster
liquid into a nonergodic gel \cite{weitzclu,bartlett}, where structural
arrest occurs. Using molecular dynamic simulations, we showed that such
structural arrest is crucially related to the formation of a long
living spanning cluster, providing evidence for the percolation nature
of the colloidal gel transition at low volume fraction and low
temperature \cite{noi_me,noi}. This scenario was confirmed by recent
experiments \cite{bartlett} and molecular dynamics
simulations\cite{sciort}, where it was shown that increasing the
volume fraction clusters coalesce into elongated structures 
eventually forming a disordered spanning network.
A realistic framework for the modelization of these systems is represented 
by DLVO interaction potentials \cite{israel}, which combine 
short-range attractions and long-range repulsions.
A suited choice of DLVO models has in fact allowed to reproduce, by means of
molecular dynamics calculations, many experimental observations, like
the cluster phase and the gel-like slow dynamics \cite{noi_me,noi,sciort}.

On the other hand, competing interactions have been studied in many other 
systems, ranging from spin systems to aqueous surfactants or mixtures 
of block copolymers, and often lead to pattern formation or to the 
creation of periodic phases
\cite{lamellar,cao,sear,muratov,reatto,marco,surf}.

In this paper, we simulate by molecular dynamics a system composed of monodisperse particles, interacting
with a short range attraction and a long range repulsion in analogy with DLVO models,
for a large range of temperatures and volume fractions. At low temperature, 
increasing the volume fraction in the region of phase space where the system 
forms a percolating network and waiting long enough, we observe that
the system spontaneously orders, to form a periodic structure composed by
parallel columns of particles.
This finding strongly suggests that the transition to the gel phase, observed 
in experiments and in numerical simulations, happens in a ``supercooled'' region, namely in a disordered phase which is metastable
with respect to crystallization. This is supported also by the results on a 
mean field model \cite{marco}.
We then study the phase diagram of the system, by evaluating the 
free energy of the disordered and ordered phases, and find the region 
where the columnar phase is stable.
We also locate the region of phase space where the stable phase is the 
lamellar one. Such phase does not form spontaneously within the observation 
times, therefore indicating a much longer nucleation time.

We have considered a system made of $1300\le N \le 2300$ particles. 
The particles interact through the effective interaction potential
\cite{noi}:
\begin{equation}
V(r)=\epsilon \left[a_1 \left(\frac{\sigma}{r}\right)^{36}
-a_2\left(\frac{\sigma}{r}\right)^6+a_3e^{-\lambda(\frac{r}{\sigma}-1)}\right],
\label{potential}
\end{equation}
where $a_1=2.3$, $a_2=6$, $a_3=3.5$, and $\lambda =2.5$.  The
potential is truncated and shifted to zero at a distance of
3.5$\sigma$. The temperature $T$ is in units of $\epsilon/k_B$, where
$k_B$ is the Boltzmann constant. The number of particles varies as we consider
different volume fractions, defined using the equivalent system of spheres
of diameter $\sigma$ as $\phi= 4/3 \pi \sigma^3 N/ L^3$ in a simulation box 
of size $L$. 
We have performed Newtonian molecular
dynamics at constant NVT using the velocity Verlet algorithm and the
Nos\'e-Hoover thermostat \cite{hoover} with time step $\Delta t=0.01
t_0$ (where $t_0=\sqrt{m\sigma^2 /\epsilon}$ and $m$ is the mass of the
particles).  For a given volume fraction $\phi$, the system is first 
thermalized at very high temperature and then quenched to
the desired temperature. We let it evolve for times ranging between
$10^5\,t_0$ and $10^6\, t_0$ ($10^7$--$10^8$ MD steps).

\begin{figure}
\begin{center}
\unitlength=5cm
\begin{picture}(2,0.8)(0,0)
\put(0,0){\mbox{\large\bf a)}}
\put(-0.02,0){\includegraphics[width=4.5cm]{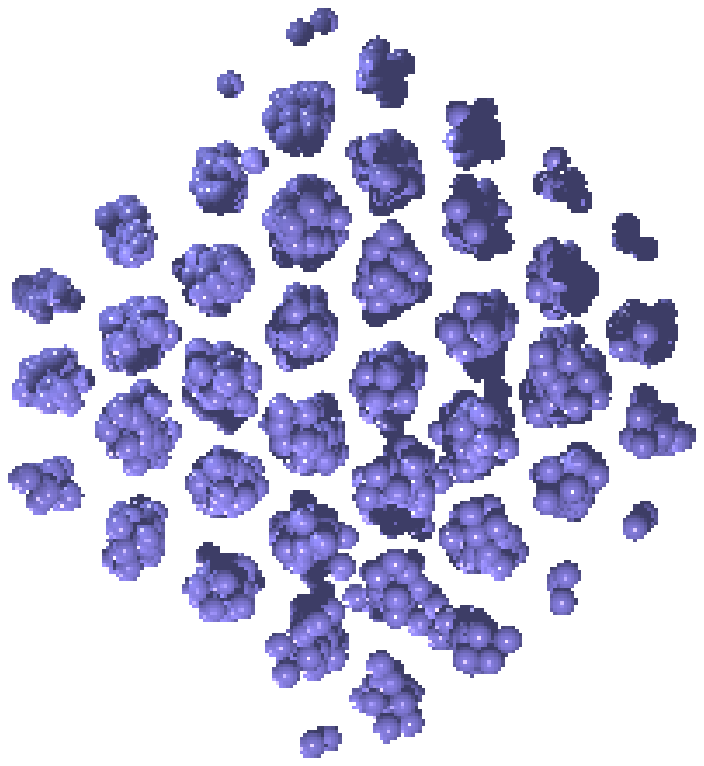}}
\put(0.9,0){\mbox{\large\bf b)}}
\put(0.95,0.1){\includegraphics[width=3.8cm]{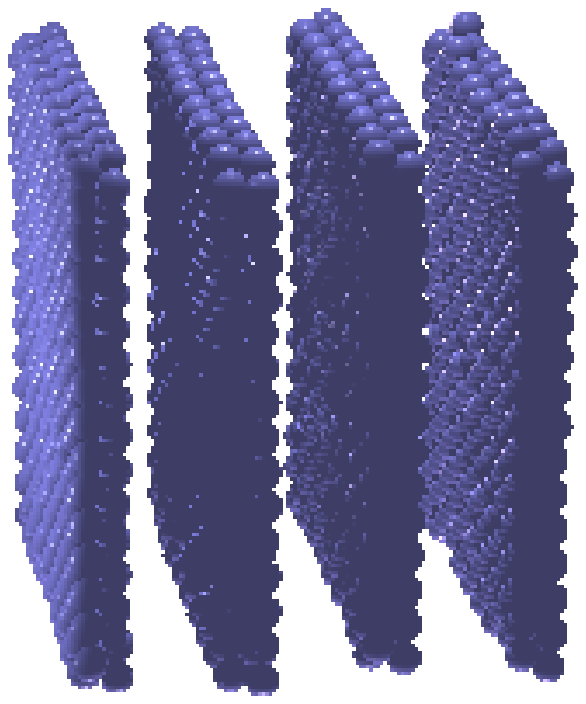}}
\end{picture}
\end{center}
\caption{a) Snapshot of one of the systems after the ordering 
(simulation time $t=10^5 t_0$),
taken from a direction parallel to the columns,
showing the two-dimensional arrangement of the axes. Since columns are not 
perfectly straight, their diameters appear to be larger than in reality; 
b) Lamellar phase at $\phi=0.26$ and $T=0.2$.}
\label{figure1}
\end{figure}

At low volume fraction ($\phi<0.2$) and high temperature ($T>0.25$),
the system remains in a disordered configuration during the simulation 
time window. Decreasing the temperature, we observe the formation of 
long living clusters with a typical size of about ten particles. 
At temperature $T=0.25$ and for volume fractions $\phi$ between $0.15$ and 
$0.23$, we find that, shortly after the quench, these clusters aggregate into 
locally rod-like structures with a length from a few to about ten molecular
diameters which finally coalesce into a disordered network. 
At longer times, between $5\times 10^4\, t_0$ and $10^6\, t_0$, these elongated
rod-like structures start to spontaneously order 
in a two-dimensional hexagonal packing of parallel columnar structures,
with a fast drop in the energy, as illustrated in Fig.\ \ref{figure1}(a).
We also observe that locally particles rearrange 
in a spiral structure which optimizes the competing interactions, very similar 
to the Bernal spiral \cite{bernal}, where each particle has six neighbors. 
Such a local structure is also indicated by experimental observations
\cite{bartlett} and found in numerical simulations with similar interactions 
\cite{sciort}.

In order to infer the phase diagram of the system at low volume
fraction, we first studied the state of the system at very low
temperature. In that limit, entropy can be neglected and the
equilibrium state of the system should be the crystalline structure
with the lowest potential energy. Of course, an exact determination 
of such a structure, 
that may depend on the volume fraction, is far from trivial. 
We have instead selected a few possible structures 
and made a comparative analysis by calculating numerically their 
potential energy as a function of the volume fraction. 
The structures considered are the following (in each
case, the lattice spacing between elements is chosen to provide the
desired volume fraction):

\noindent
1) Cluster crystal: Three-dimensional hexagonal close packing of
nearly spherical clusters, each cluster being composed of thirteen
particles (a central one in contact with twelve external ones).  In
this case particles have 5.54 neighbors on the average. These
clusters were chosen because of their similarity to the long-living
clusters observed at low temperature and volume fraction, 
composed generally by one particle surrounded by eight to
eleven neighbors.

\noindent
2) Columnar phase: Two-dimensional hexagonal packing of parallel Bernal
spirals. In this case, each particle has six neighbors.

\noindent
3) Lamellar phase: Parallel planes, each plane being formed by one or more 
layers of hexagonally packed particles (Fig.\ \ref{figure1}(b)). 
In this case, the average
number of neighbors per particle is between six (one layer) and twelve
(when the number of layers increases the structure becomes an hexagonal
close-packed lattice of particles).

\begin{figure}
\begin{center}
\includegraphics[width=7cm]{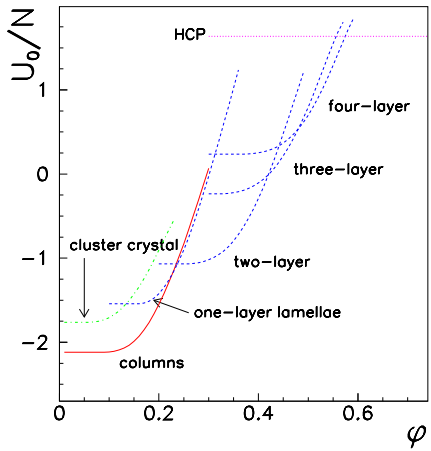}
\end{center}
\caption{Minimum of the total potential energy per particle, $U_0/N$,
for several structures as a function of the volume fraction.}
\label{figure2}
\end{figure}

After having placed the particles to form these structures with the
distance between nearest neighbors corresponding to the minimum of the
pair-potential, we have calculated the nearest minimum of the potential
energy landscape, by minimizing the total potential energy
$U(\r_1,\ldots,\r_N)$ as a function of the particle positions, using
a conjugate-gradient algorithm \cite{nrc}.  The potential energy per particle 
$U_0/N$ of the minimum is shown in Fig.\ref{figure2}.

The cluster crystal never provides the lowest energy because of the
limited number of neighbors per particle. For volume fractions
$\phi<0.235$, the columnar phase of Bernal spirals is energetically preferred.
Indeed, in this case the combination of a relatively high number of first 
neighbors and a low number of second and third neighbors minimizes the 
repulsive part of the interaction potential. Increasing the volume fraction, 
the columns get closer and the repulsion between them becomes relevant. At
$\phi>0.235$, the potential energy of the columnar phase becomes higher than
that of the two-layer lamellae. In fact, since lamellae are more distant 
than columns for a given volume fraction, the repulsion between lamellae
is weaker. It is interesting to note that one-layer lamellae are never
energetically preferred and that their minimum potential energy
becomes close to the one of the columns at high volume fraction. At
$\phi>0.42$, the three-layer lamellae exhibit the lowest potential
energy, and so on. Similar results were recently found in a $2d$ DLVO
system \cite{reatto}.

\begin{figure}
\begin{center}
\includegraphics[width=7cm]{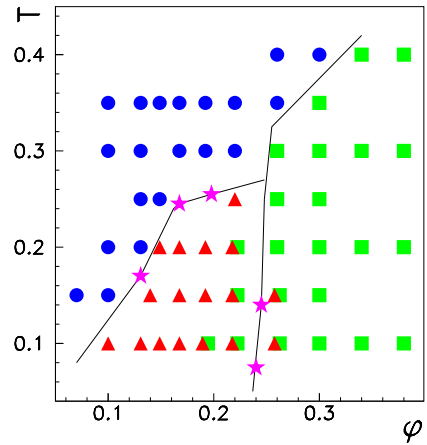}
\end{center}
\caption{$T$-$\phi$ phase diagram of the system. Circles represent
the disordered phase, triangles the columnar phase and squares the
lamellar phase. Stars represent points where the calculated free energy of the two phases crosses.
Solid lines are a guide for the eyes.}
\label{figure3}
\end{figure}

In order to test the stability of these structures as the temperature
increases, we performe the following experiment. Given a volume
fraction and a temperature, we let evolve a crystalline configuration of 
columns or two-layer lamellae by means of molecular dynamics at constant 
temperature for $10^6t_{0}$. The obtained phase diagram is shown in 
Fig.\ \ref{figure3}. Circles represent state points where both columns
and lamellae break down before the end of the simulation (disordered phase).  
Triangles are the points where lamellae break down, but columns are stable 
(columnar phase). Squares are the points where lamellae are stable and columns
break down (lamellar phase). Finally, the points where triangles and squares 
overlap correspond to the states where both columns and lamellae remain stable 
until $t=10^6t_0$. We note that 
near the boundaries with the disordered phase, columns or lamellae
become fuzzy, but the modulation of the density is still clearly visible.

To check the results found by molecular dynamics simulations and to
better estimate the phase boundaries, we have computed the free energy
of the three phases that seem to be relevant for $\phi<0.4$: disordered
phase, columnar phase, and two-layer lamellar phase.
The free energy of the disordered phase is computed by thermodynamic
integration \cite{thermo} along the following path: from the perfect gas limit
($V\to\infty$) down to the desired volume $V$ along the isotherm
$T_0=1$. The free energy at $(V,T_0)$ is therefore given by
\be
\frac{F_0}{NT_0}=\ln\left(\frac{N}{V}\right)
+\frac{3}{2}\ln\left(\frac{2\pi\hbar^2}{mT_0}\right)-1
+\int\limits_V^\infty\frac{\Delta P(V^\prime)}{NT_0}\,dV^\prime
\ee
where $\Delta P(V)=P(V)-\frac{NT_0}{V}$ is the excess pressure with
respect to the perfect gas. Since values of $\hbar$ and the mass $m$ merely
shift the free energy by a constant, we set $\hbar=1$ and $m=1$.
We have then integrated along the isochore $V$, from $T_0$ to the desired
temperature, and obtained
\be
\frac{F}{NT}=\frac{F_0}{NT_0}
+\frac{3}{2}\ln\left(\frac{T_0}{T}\right)+\int\limits_T^{T_0}\frac{U(T^\prime)}%
{N{T^\prime}^2}\,dT^\prime.
\ee
Starting from $T=0$, we have calculated the free energy of the two crystalline 
phases by integrating along an isochore. At very low temperatures the potential 
energy landscape can be approximated by a parabolic function
\be
U=U_0+\frac{1}{2}\sum_{i=1}^{3N}\lambda_i x_i^2,
\ee
where $\lambda_i$ are the eigenvalues of the Hessian matrix at the
minimum of the potential. The free energy is then calculated as
\be
\frac{F}{NT}=\frac{U_0}{NT}+
\frac{1}{2}\sum_{i=1}^{3N}\ln\lambda_i+\frac{3}{2}\ln\left(\frac{\hbar^2}%
{mT^2}\right)-\int\limits_0^T\frac{\Delta U(T^\prime)}{N{T^\prime}^2}\,dT^\prime,
\ee
where $\Delta U(T)=U(T)-U_0-\frac{3NT}{2}$.

\begin{figure}
\begin{center}
\includegraphics[width=7cm]{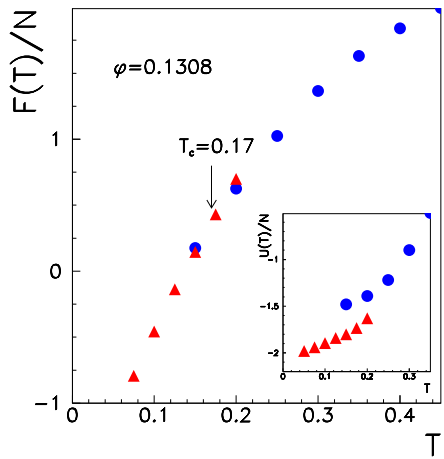}
\end{center}
\caption{Free energy per particle $F(T)/N$ of the disordered phase (circles) and
of the columnar phase (triangles) as a function of temperature at
$\phi=0.1308$. Inset: potential energy per particle $U(T)/N$.}
\label{figure4}
\end{figure}

In Fig.\ \ref{figure4} the free energy of the disordered phase and of 
the columnar phase at $\phi=0.1308$ are shown as a function of the
temperature.  The curves cross at the temperature ($T_c=0.17$) 
corresponding to the first order transition from the columnar phase to
the disordered phase. This, together with the other transition points, are
marked in Fig.\ \ref{figure3} by a star \cite{nota2}.

These results suggest the following scenario for colloidal gelation at low 
volume fraction and low temperature. Just remind that in real colloidal 
systems the control parameter corresponding to low temperature is a high
effective attraction strength. The competition between 
attractive and repulsive interaction induces a
typical modulation length clearly detected at very low volume fraction
in the cluster phase. 
As the volume fraction is increased, clusters
are prone to aggregate in spite of the long-range repulsion. This
results in the growing of elongated structures which keep track of the
modulation length. Eventually, a modulated phase forms,
with a columnar geometry or a lamellar one at higher volume fraction. Of
course the precursors of the modulated structures will have defects
such as local inhomogeneities and branching points. Furthermore, the
slow dynamics due to the viscosity of the solvent and to the imperfect
shape of the particles may also hinder the ordered phase, producing
long-living metastable disordered states.
Hence, when the volume fraction is high enough, an interconnected
spanning structure is formed and a gel-like behavior is observed as it
is frequently the case in the experiments.

The authors thank Gilles Tarjus for interesting discussions and for
pointing out the existence of lamellar phases in %
systems with competing interactions. 
The research is supported by 
MIUR-PRIN 2004, MIUR-FIRB 2001, CRdC-AMRA,  
the Marie Curie Reintegration Grant MERG-CT-2004-012867 and 
EU Network Number MRTN-CT-2003-504712.

\begin{thebibliography}{99}
\bibitem{nota2} In principle, the first order transition line
$T-\phi$ can be calculated also with the Clausius-Clapeyron equation
$\frac{dT}{dV}=-\frac{T\Delta P}{\Delta U}$, where $\Delta P$ and
$\Delta U$ are the pression and energy differences between the two
phases.  However, in this way the result depends entirely on
quantities evaluated at the transition, where the fluctuations and
hence the errors are large.
\end{thebibliography}
\end{document}